# A PRELIMINARY STUDY OF OXIDATION OF LIGNIN FROM RUBBER WOOD TO VANILLIN IN IONIC LIQUID MEDIUM

A. A. SHAMSURI[a]*, D. K. ABDULLAH[b]

[a]*Laboratory of Molecular Biomedicine, Institute of Bioscience, University Putra Malaysia, 43 400 UPM Serdang, Selangor, Malaysia*
*E-mail*: adlie@putra.upm.edu.my
[b]*Department of Chemistry, Faculty of Science, University Putra Malaysia, 43 400 UPM Serdang, Selangor, Malaysia*

ABSTRACT

In this study, lignin was oxidised to vanillin by means of oxygen in ionic liquid (1,3-dimethylimidazolium methylsulphate) medium. The parameters of the oxidation reaction that have been investigated were the following: concentration of oxygen (5, 10, 15 and 20 ft$^3$ h$^{-1}$), reaction time (2, 4, 6, 8 and 10 h) and reaction temperature (25, 40, 60, 80 and 100°C). The Fourier transform infrared spectroscopy, high performance liquid chromatography and ultraviolet-visible analyses were used to characterise the product. The results revealed vanillin as the product obtained via the oxidation reaction. The optimum parameters of vanillin production were 20 ft$^3$ h$^{-1}$ of oxygen for 10 h at 100ºC. In conclusion, 1,3-dimethylimidazolium methylsulphate could be used as an oxidation reaction medium for the production of vanillin from rubber wood lignin.

AIMS AND BACKGROUND

Lignin is second in abundance component of wood after cellulose that can be potential substrates for the production of vanillin. Vanillin has been used as flavouring materials and aroma in food industry and also in the fragrance industry for instance perfumes, hygienic products and to cover up disagreeable scent or taste in drugs as well. It can also be used in chemical intermediary for the production of pharmaceuticals and other fine chemicals[1,2]. The presence of small quantity of vanillin in vanilla beans or seed pods of the tropical *Vanilla* orchid makes its extraction limited and not economically feasible[3]. Thus, the oxidation of lignin to produce vanillin has been investigated extensively[4,5]. In addition, lignin-based vanillin is suitable as flavouring because it has a richer flavour profile than oil-based flavouring[1].

---

* For correspondence.

Lignin can be isolated from lignocellulosic materials by using the Kraft pulping or sulphate process. The process involved the use of concentrated and strong bases, specifically a mixture of sodium hydroxide and sodium sulphide, known as white liquor[6]. Black liquor that contains lignin fragments after processing of lignocellulosic materials by this isolation process can be used to produce vanillin via oxidation reaction[7]. However, additional steps are needed after the oxidation reaction to neutralise the high pH value of black liquor in relation to the vanillin recovery. Besides, this process to produce vanillin is no longer accepted because of corrosivity, hazardous and can cause harm to the environment[8]. Another method has been developed to isolate lignin is organosolvent process[9]. Nevertheless, this process also has its own drawbacks due to the use of organic solvents that are highly volatile, highly flammable and toxic. Moreover, the process also usually requires high pressures conditions[10] and the state-of-the-art equipments are often necessitated for oxidation of lignin since low boiling point of organic solvents that were used as a medium have high vapour pressure.

Ionic liquids are considered as green solvents because they indicate some interesting properties such as negligible vapour pressure (being non-volatile), environmentally-benign, non-toxic, recyclable, non-flammable and chemically inert[11–14]. Ionic liquids are capable to dissolve and blend biopolymers with high efficiency, no severe side-reaction occurred and simplicity in recovery of the products[15,16]. Since ionic liquids offer a potentially clean medium for carrying out chemical reactions or processes, more attention has been paid on lignocellulosic materials, for example cellulose and lignin[17,18]. Up to now, the oxidation studies of lignin in ionic liquids medium are less reported, especially by using hydrophilic ionic liquid and in the absence of any catalyst[19]. In this preliminary study, rubber wood was used as a raw material. The lignin was isolated by means of ionic liquid from rubber wood, the isolated lignin was oxidised to produce vanillin. The process was carried out in hydrophilic ionic liquid medium and no catalyst required for avoiding possible metal-contamination of the product. The product was characterised by the Fourier transform infrared spectroscopy, high performance liquid chromatography and ultraviolet-visible analyses. The effect of experimental reaction conditions on the optimisation of vanillin production was also studied.

EXPERIMENTAL

MATERIALS

1,3-Dimethylimidazolium methylsulphate and methanol were purchased from Merck. Rubber wood (*Hevea brasiliensis*) used in the experiments was supplied by local sawmill and wood processing factory in Malaysia. Oxygen gas was obtained from Mox-Linde Gases. Chloroform was procured from Fisher Scientific. All chemicals were used as received.

ISOLATION OF LIGNIN FROM RUBBER WOOD

The isolation of lignin from rubber wood by using ionic liquid was done according to a procedure described elsewhere[12]. Rubber wood was grinded, sieved and kept in the oven to remove moisture. Then, 52.50 g of rubber wood (oven-dry weight) were placed in a glass flask, 5.0 mol of 1,3-dimethylimidazolium methylsulphate were poured into a glass flask. The flask was stirred and heated by placing in oil bath at 100°C for 2 h. After the flask was cooled to the room temperature, the insoluble component in the solution was then separated by filtration under reduced pressure. Then, the soluble lignin was separated from ionic liquid through precipitation with methanol. The isolated lignin was filtered off and washed thoroughly with distilled water for several times. They were collected and dried into a vacuum oven at 85°C for 24 h. The isolated lignin was kept over a desiccant before further oxidation.

OXIDATION OF LIGNIN TO VANILLIN

2.0 g of the isolated lignin were dissolved into 18.0 g of 1,3-dimethylimidazolium methylsulphate in a beaker with stirring to prepare lignin solution with concentration of 10 wt.%. The solution was transferred into a round-bottom flask and then the flask was mounted on a rotary evaporator apparatus equipped with oxygen tank for the oxidation reaction. After specific reaction condition, the oxidative mixture was mixed with 50 g of chloroform with stirring in a beaker for 30 min followed by filtration. The filtrate was then transferred to a separation funnel in order to extract the product and separate any lignin solution. The solution at the bottom of the funnel was collected. The product soluble in chloroform was used in the experiments for characterisation. The parameters investigated include the concentration of oxygen (5, 10, 15 and 20 ft$^3$ h$^{-1}$), reaction time (2, 4, 6, 8 and 10 h) and reaction temperature (25, 40, 60, 80 and 100°C). The experiments were performed in triplicate.

CHARACTERISATION

*Fourier transform infrared* (*FTIR*). FTIR analyses were carried out by using a Perkin Elmer Spectrum 100 Series spectrometer to determine the presence of vanillin functional groups in the product after the oxidation. The FTIR spectra were obtained by using a universal attenuated total reflectance (UATR) equipped with a ZnSe-diamond composite crystal accessory. The spectra resolution was 4 cm$^{-1}$, the scanning wavenumber ranged from 4000 to 800 cm$^{-1}$ and each spectrum was 16 scans. All spectra were rationed against the reference spectrum of background.

*High performance liquid chromatography* (*HPLC*). The identification of the product was carried out based on qualitative analysis in an Agilent 1200 Series HPLC and a column (5-pm octadecasilane silica, Nucleosil-C18, 150 × 4.6 mm). The separation was obtained using a linear gradient of 2 solvent systems: solvent A (water + acetic acid, 94:6) and solvent B (methanol + acetonitrile + acetic acid, 95:4:1). A linear gradient was run for over 30 min from 0 to 40% B as eluent at a flow rate of 1 ml min$^{-1}$.

The product was detected with a DAD detector at 280 nm by computer comparison of the retention times and peak areas with the standard vanillin.

*Ultraviolet-visible* (*UV-vis.*) *spectroscopy*. UV-vis. spectra were measured on a Perkin Elmer Lambda 25 UV-vis. spectrometer in the wavelength range of between 240 and 330 nm with scan speed of 960 nm min$^{-1}$. Lamp change wavelength was 355 nm and slit width 1.00 nm. A 2.5 ml aliquot of the product was placed directly into the quartz cuvette where the beam could directly pass through it to measure the absorbance (*A*).

RESULTS AND DISCUSSION

FTIR CHARACTERISATION

Figure 1 displays the FTIR spectra of the standard vanillin and the product, the bands of spectra are summarised in Table 1 and the chemical structure of vanillin is shown in Fig. 2. The functional groups and corresponding molecular motions are determined based on the specific band. As can be seen in Fig 1, the FTIR spectra of the standard vanillin (*a*) and the product (*b*) exhibited strong intensity and broad bands at 3475.26 and 3476.48 cm$^{-1}$, respectively. These bands could be assigned to the O–H bond stretching of the phenol group, while the bands with medium intensity at 3021.48 and 3022.39 cm$^{-1}$ are corresponded with the C–H bond stretching of the aromatic ring group. The characteristic C=O bond stretching of the aldehyde group results from medium intensity of the bands at 1644.32 and 1645.27 cm$^{-1}$. The bands with strong intensity that are caused by C–O bond stretching of the ether group belong to 1214.25 and 1215.42 cm$^{-1}$. The strong intensity of the bands at 1001.27 and 1002.15 cm$^{-1}$ are associated with C–O bond stretching of the phenol group. From Table 1 the spectrum bands of the standard vanillin and the product illustrated the equality between them and the presence of the functional groups of vanillin (Fig. 2) are explicitly observed in the product. Hence, the product from the oxidation reaction could be recognised as a vanillin. However, HPLC analyses were additionally conducted to confirm vanillin as the product of oxidation.

**Table 1**. FTIR bands of the standard vanillin and the oxidation product

| Standard vanillin (cm$^{-1}$) | Product (cm$^{-1}$) | Band origin | Functional group |
|---|---|---|---|
| 3475.26 | 3476.48 | O–H stretching | phenol |
| 3021.48 | 3022.39 | C–H stretching | aromatic ring |
| 1644.32 | 1645.27 | C=O stretching | aldehyde |
| 1214.25 | 1215.42 | C–O stretching | ether |
| 1001.27 | 1002.15 | C–O stretching | phenol |

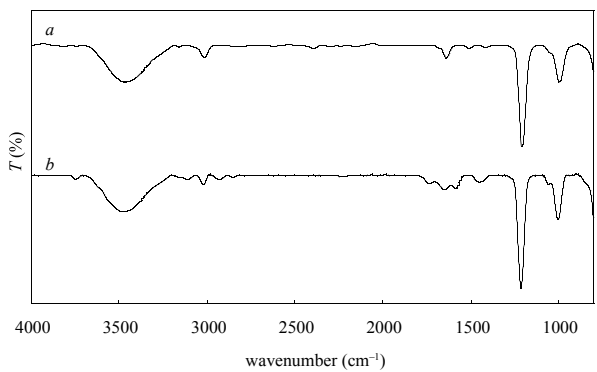

**Fig. 1**. FTIR spectra of the standard vanillin (*a*) and the oxidation product (*b*)

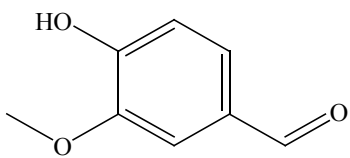

**Fig. 2**. Chemical structure of vanillin

HPLC CHARACTERISATION

Figures 3 and 4 present the HPLC chromatograms of the standard vanillin and the product, respectively. The chromatograms showed a major peak for the retention time of the product (11.312 min) which is nearly the same as the retention time of the standard vanillin (11.318 min). This observation corresponds to the result that was obtained from the FTIR spectra. This qualitative analysis also indicated there was no significant difference in the chromatogram of the product from the oxidation reaction with the standard vanillin. However, the unanticipated high signal/noise ratio of the product believed to be due to the existence of small amount of impurities. Based on the analysis, the peaks of impurities did not show any useful information regarding other compounds because their concentrations are relatively low, thus, they are not considered further. Nevertheless, UV-vis. measurements were conducted to prove the presence of insignificant amounts of impurities in the product.

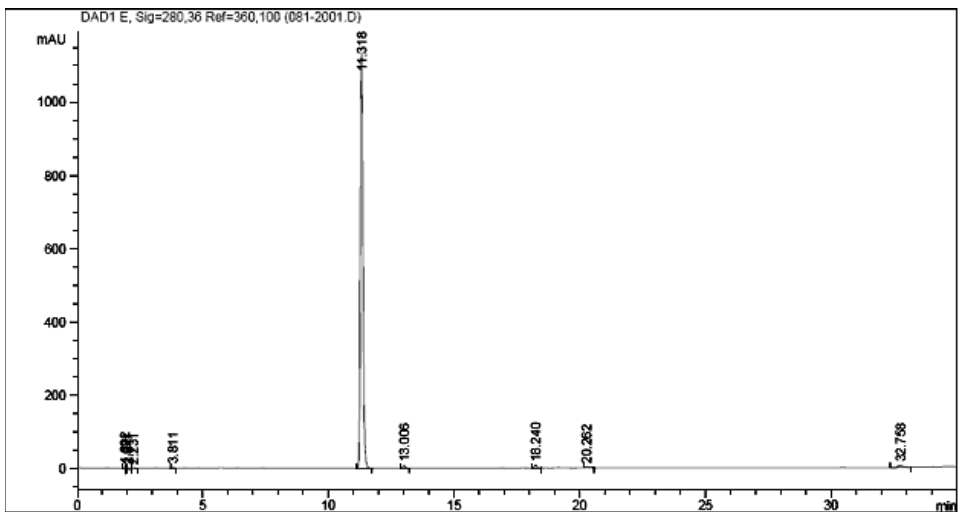

**Fig. 3**. HPLC chromatogram of the standard vanillin

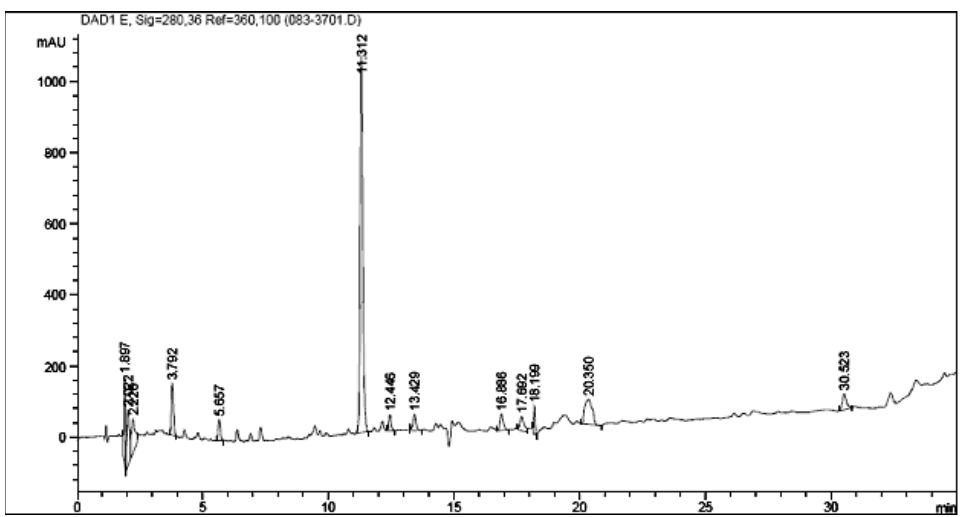

**Fig. 4**. HPLC chromatogram of the oxidation product

UV-vis. CHARACTERISATION

Figure 5 shows the UV-vis. spectra of the standard vanillin and the product. The absorption peaks for both spectra are almost similar especially for the peaks in the region from 242 to 254 nm. However, the spectrum of the product contained some small absorption peaks arising between 284 and 302 nm in comparison with the standard vanillin. These results suggested that the insignificant absorption peaks could be attributed to the amount of impurities that are present in the product in insignificant amounts. These peaks could be assumed negligible as long as the main broad peaks (242 to 254 nm) are not changed to sharp and narrow peaks, this is most likely the main broad peaks are far more significant compared to the other peaks. The UV-vis. results are entirely consistent with the results obtained from the HPLC chromatograms.

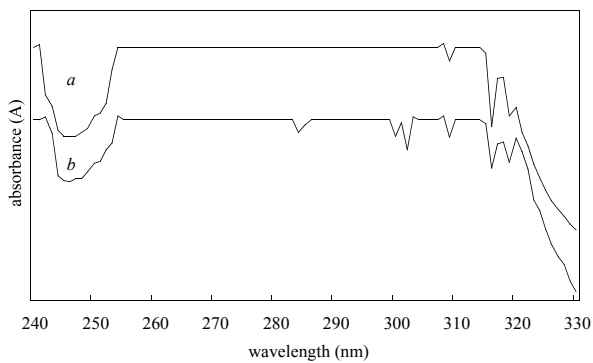

**Fig. 5**. UV-vis. spectra of the standard vanillin (*a*) and the oxidation product (*b*)

OPTIMUM CONDITIONS FOR OXIDATION OF LIGNIN TO VANILLIN

The product that has been characterised above is obtained under optimum reaction conditions. Lignin undergoes depolymerisation during the oxidation reaction; as a result, the rate of depolymerisation depends on the oxidation conditions. The primary chemical structure of lignin does not changed during the dissolution in ionic liquid. However, it changes along with the formation of vanillin when the concentration of oxygen is greater. The reaction time of the oxidative mixture has to be maximised to extend the depolymerisation. The temperature has also indicated similar effect on oxidation reaction, whereby high temperature increases the solubility of oxygen in the ionic liquid and enhances the lignin depolymerisation rate. Therefore, the optimum parameters of vanillin production were 20 ft$^3$ h$^{-1}$ of oxygen for 10 h at 100ºC that led to lignin depolymerisation with high efficiency. In contrast, increase in ionic liquid concentration does not result in improvement of the vanillin production for all investigated parameters, the resultant product is almost the same at the end of the oxidation reaction. On the other hand, this preliminary study also showed that the product acquired from these optimum reaction conditions released a scent identical to the standard vanillin.

CONCLUSIONS

The FTIR spectroscopy results indicated the presence of vanillin functional groups in the product. The HPLC results revealed that the retention time of the product is nearly the same as that of the standard vanillin. UV-vis. spectra demonstrated absorption peak of the product very similar to those of the standard vanillin. The optimum parameters of vanillin production were 20 ft$^3$ h$^{-1}$ of oxygen for 10 h at 100ºC. In conclusion, the vanillin has been successfully obtained by means of rubber wood-based lignin via oxidation reaction in an ionic liquid (1,3-dimethylimidazolium methylsulphate) medium.

ACKNOWLEDGEMENT

The authors gratefully acknowledge the technical support and the facilities provided by the Institute of Bioscience (IBS), University Putra Malaysia (UPM). Research for this paper was supported by UPM under the Research University Grant Scheme (05-01-09-0617RU). This paper is dedicated to Prof. Dr. Dzulkefly Kuang Abdullah, the last Head of Laboratory of Industrial Biotechnology, IBS, UPM. We also thank Prof. Dr. Slavi K. Ivanov, Mrs. M. Boneva and anonymous referees for valuable comments on this paper.

REFERENCES

1. L. J. ESPOSITO, K. FORMANEK, G. KIENTZ, F. MAUGER, V. MAUREAUX, G. ROBERT, F. TRUCHET (Eds): Vanillin. Vol. 24: Kirk-Othmer Encyclopedia of Chemical Technology. 4th ed. John Wiley & Sons, New York, 1997, p. 812.
2. M. J. W. DIGNUM, J. KERLER, R. VERPOORTE: Vanilla Production: Technological, Chemical, and Biosynthetic Aspects. Food Rev. Int., **17**, 119 (2001).
3. N. J. WALTON, M. J. MAYER, A. NARBAD: Vanillin. Phytochemistry, **63**, 505 (2003).
4. M. ZABKOVA, E. A. des SILVA, A. E. RODRIGUES: Recovery of Vanillin from Lignin/Vanillin Mixture by Using Tubular Ceramic Ultrafiltration Membranes. J. Membr. Sci., **301**, 221 (2007).
5. P. SRIDHAR, J. D. ARAUJO, A. E. RODRIGUES: Modeling of Vanillin Production in a Structured Bubble Column Reactor. Catal. Today, **105**, 574 (2005).
6. O. WALLBERG, A-S. JÖNSSON: Separation of Lignin in Kraft Cooking Liquor from a Continuous Digester by Ultrafiltration at Temperatures above 100°C. Desalination, **195**, 187 (2006).
7. C. FARGUES, A. MATHIAS, A. RODRIGUES: Kinetics of Vanillin Production from Kraft Lignin Oxidation. Ind. Eng. Chem. Res., **35**, 28 (1996).
8. M. B. HOCKING: Vanillin: Synthetic Flavoring from Spent Sulfite Liquor. J. Chem. Educ., **74**, 1055 (1997).
9. J. I. BOTELLO, M. A. GILARRANZ, F. RODRIGUEZ, M. OLIET: Preliminary Study on Products Distribution in Alcohol Pulping of *Eucalyptus globulus*. J. Chem. Technol. Biotechnol., **74**, 141 (1999).
10. L. BARBERA, M. A. PELACH, I. PEREZ, J. PUIG, P. MUTJE: Upgrading of Hemp Core for Papermaking Purposes by Means of Organosolvent Process. Ind. Crop. Prod., **34**, 865 (2011).
11. A. A. SHAMSURI, D. K. ABDULLAH: Protonation and Complexation Approaches for Production of Protic Eutectic Ionic Liquids. J. Phys. Sci., **21**, 15 (2010).
12. A. A. SHAMSURI, D. K. ABDULLAH: Isolation and Characterization of Lignin from Rubber Wood in Ionic Liquid Medium. Mod. Appl. Sci., **4**, 19 (2010).
13. A. A. SHAMSURI, D. K. ABDULLAH: Synthesizing of Ionic Liquids from Different Chemical Reactions. Singap. J. Sci. Res., **1**, 246 (2011).

14. A. A. SHAMSURI, R. DAIK: Plasticising Effect of Choline Chloride/Urea Eutectic-based Ionic Liquid on Physicochemical Properties of Agarose Films. BioResources, **7**, 4760 (2012).
15. A. A. SHAMSURI, D. K. ABDULLAH, R. DAIK: Fabrication of Agar/Biopolymer Blend Aerogels in Ionic Liquid and Co-solvent Mixture. Cell Chem. Technol., **46**, 45 (2012).
16. H. XIE, S. LI, S. ZHANG: Ionic Liquids as Novel Solvents for the Dissolution and Blending of Wool Keratin Fibers. Green Chem., **7**, 606 (2005).
17. L. FENG, Z.-I. CHEN: Research Progress on Dissolution and Functional Modification of Cellulose in Ionic Liquids. J. Mol. Liq., **142**, 1 (2008).
18. Y. PU, N. JIANG, A. J. RAGAUSKAS: Ionic Liquid as a Green Solvent for Lignin. J. Wood Chem. Technol., **27**, 23 (2007).
19. K. STÄRK, N. TACCARDI, A. BÖSMANN, P. WASSERSCHEID: Oxidative Depolymerization of Lignin in Ionic Liquids. ChemSusChem., **3**, 719 (2010).